\documentclass{article}

\usepackage{PRIMEarxiv}

\usepackage[utf8]{inputenc} 
\usepackage[T1]{fontenc}    
\usepackage{hyperref}       
\usepackage{url}            
\usepackage{booktabs}       
\usepackage{amsfonts}       
\usepackage{nicefrac}       
\usepackage{microtype}      
\usepackage{lipsum}
\usepackage{fancyhdr}       
\usepackage{graphicx}       
\graphicspath{{media/}}     
\usepackage{multirow} 
\usepackage{makecell}
\usepackage{float}
\usepackage{longtable}
\title{Temporal Importance Factor for Loss Functions for CTR Prediction}

\author{
  Ramazan Tarık Türksoy \\
  Huawei Turkey R\&D Center \\
  Bogazici University \\
  Istanbul\\
  \texttt{ramazan.tarik.turksoy1@huawei.com} \\
     \And
  Beyza Türkmen \\
  Huawei Turkey R\&D Center \\
  Istanbul\\
  \texttt{beyza.turkmen1@huawei.com} \\
     \And
  Furkan Durmuş \\
  Huawei Turkey R\&D Center \\
  Bogazici University \\
  Istanbul \\
  \texttt{furkan.durmus1@huawei.com} \\
}

\begin{document}
\maketitle

\begin{abstract}
Click-through rate (CTR) prediction is an important task for the companies to recommend products which better match user preferences. User behavior in digital advertising is dynamic and changes over time. It is crucial for the companies to capture the most recent trends to provide more accurate recommendations for users. In CTR prediction, most models use binary cross-entropy loss function. However, it does not focus on the data distribution shifts occurring over time. To address this problem, we propose a factor for the loss functions by utilizing the sequential nature of user-item interactions. This approach aims to focus on the most recent samples by penalizing them more through the loss function without forgetting the long-term information. Our solution is model-agnostic, and the temporal importance factor can be used with different loss functions. Offline experiments in both public and company datasets show that the temporal importance factor for loss functions outperforms the baseline loss functions considered.
\end{abstract}


\section{Introduction}

Predicting click-through rate (CTR) is critical to maximize advertising revenue for companies by keeping user-item interaction at an optimal level. The ability to successfully model the user-item interactions is directly related with the revenue that can be obtained \cite{Richardson2007, Wang2020a}. CVR prediction is another recommendation task which aims the user who clicked on the ad to take the desired action, e.g., purchase the current product. It can improve the user experience, and increase the company revenues. Therefore, there are several mechanisms for modeling the interactions between users and items. Traditional recommender systems perform user-item matching via linear interactions and may be inadequate at modeling user-item relationships, which are actually much more complex and dynamic \cite{Wolfe2010}. Low-order interactions arise from the linear relationships between the user and the item such as the frequency of purchase of a product. On the other hand, high-order interactions come from the relationships of features with each other at different levels, rather than just looking at linear relationships \cite{Zhao2020a}. These can be mentioned as the time interval a user views a product, browsing history, and relationship with other users. Complex and deep models used in the industry obtain good results, since they can capture high-order interactions. 

Recent studies have developed various strategies to capture feature interactions using deep CTR models \cite{masknet, deepfm, autoint, DCN}. Nevertheless, the nature of the data varies across the application domains of CTR prediction. Therefore, not every model can successfully capture all of the feature interactions successfully in every area. Different training strategies and methods, such as different regularization methods, interaction layers, and loss functions, can also improve the models in capturing interactions between features \cite{YiPeng2023, Rendle2012, 10.1145/3447548.3467208, Himan2019}. In this way, robust and effective recommender systems can be built.

The main motivation of this study is to reduce the effect of rapidly changing user preferences, called concept drift \cite{liu2023concept, 10.5555/2887770.2887854}, on the success of the model and to ensure that users are shown the most appropriate items. In recommender systems, it is common to collect long-term data to make user-item interactions generalizable. For example, during the exam period, someone might focus on course content online, while after the exam week, they might browse content related to games or movies. When this person's data is collected, the user's browsing history has changed significantly in a week, and the deep learning model will learn this person along with the course and game contents. The collected data inevitably contains a temporal flow. As this flow changes the data statistics, giving each sample the same importance will diminish the success of the model in production.

There are studies on concept drift by considering the dynamic nature of user preferences. Koychev et al. weights the samples according to the viewing frequency at a given time and applies the forgetting mechanism to the old-time samples. The specified time interval is set as the window size parameter \cite{Koychev2004}. In their study, Li et al. attempted to model the drift of item popularity and user preferences over time using the Time-Interval Attention Layer \cite{Li2020a}. Lathia et al. investigated the effects of temporal diversity on recommender systems in their studies. Their findings show that spreading diversity over a large temporal range allows users to interact more \cite{Lathia2010}. While there are studies that consider temporal importance, most of the methods used in the industry build models without considering temporal information. Even training the models is done by shuffling the data.

\begin{figure}[h]
  \centering
  \includegraphics[scale=0.13]{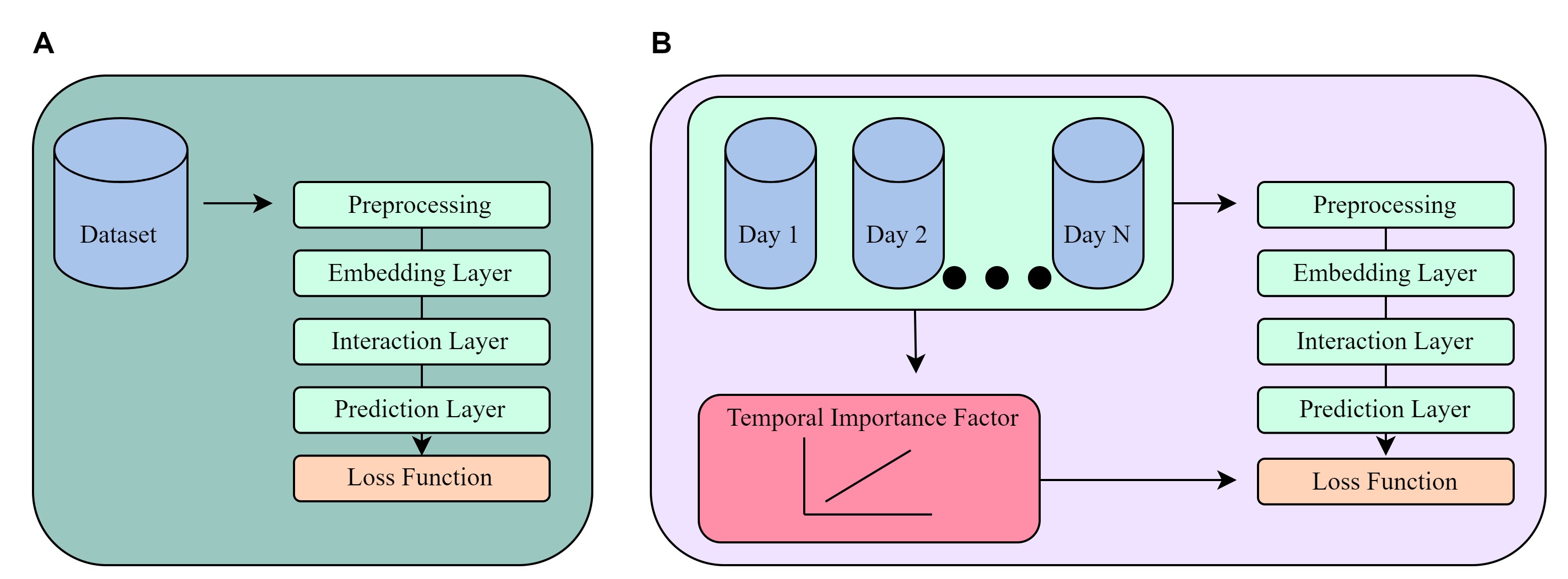}
  \caption{The left side represents the traditional model training using a loss function. The right side shows training with the proposed temporal importance factor. On the left side, the entire training set is preprocessed directly and the model is trained without temporal information. On the right side, the dataset is split by days and the samples from these splits are sent to the temporal importance factor and weighted accordingly.}
  \label{mainFigure}
\end{figure}

In this paper, we address the above challenges and propose a Temporal Importance Factor in loss functions. Instead of training the model with shuffled data, we assign temporal importance values to each sample and weight them. Instead of completely deleting the old data, the information from old samples is not completely lost. By giving more weight to samples from the recent days, we enable the model to obtain more information from recent user-item interactions. The weights are given to the chronologically ordered samples during preprocessing. These weights are added to the loss function during the model training. More recent samples are penalized more by giving higher weights during loss function optimization. The main scheme of the proposed method is shown in Figure \ref{mainFigure}. Considering the conducted extensive experiments, the developed method has proven its effectiveness by increasing performance metrics in both public and company datasets.

The main contributions of this paper are listed below:
\begin{itemize}
\item Dynamic user behaviors are better modeled to provide a more robust user-item interaction against concept drifts.
\item The developed method is model-agnostic.
\item The developed method has proven its effectiveness by improving CTR prediction performance in both public and company datasets.
\end{itemize}

\section{Related Work}
In this section, we briefly review existing models for CTR prediction. We also discuss how these models utilize temporal information of the data to achieve better performance, and the importance of the loss function in CTR.

\subsection{Deep CTR Models}
Most of the recent CTR models focus on learning sophisticated interactions between different features, considering each sample of user-item interactions individually. Representative models include Wide\&Deep \cite{wideAndDeep}, DeepFM \cite{deepfm}, and DCN \cite{DCN}. Wide\&Deep \cite{wideAndDeep} has a parallel architecture including wide linear part and a deep neural network (DNN) part to learn both explicit and implicit feature interactions. This dual structure is useful in terms of memorization and generalization. However, the wide part requires manual effort to model the low-order interactions. DeepFM \cite{deepfm} proposes the FM layer to learn the second-order feature interactions from the inner products of the vectors instead of the wide part to avoid this manual effort. Moreover, DCN \cite{DCN} uses a cross layer to replace the wide part and improve the efficiency of the DNN model.

The feature interaction layer is one of the main structures of deep CTR models. There are many innovative studies to improve the capture of feature interactions. The attention mechanism has been shown to be effective for interaction layers in CTR prediction models such as AutoInt \cite{autoint} and DIN \cite{DIN}. AutoInt \cite{autoint} uses a multi-head self-attention network to model feature interactions in low-dimensional space. DIN \cite{DIN}, is an example of user interest modeling and captures the various interests of users by exploiting historical behaviors.

\subsection{Approaches Utilizing Temporal Information in CTR Models}
Although many application domains record interactions between users and objects over time, most studies do not use the temporal relationship between interactions. A number of previous works show that temporal information can be used to build more robust user models and capture additional behavioral patterns. Unlike DIN \cite{DIN}, DIEN \cite{dien} model accounts for the changing trend of the interest using temporal information on the data. DIEN \cite{dien} proposes an auxiliary loss, which can better address the problem of interest drifting problem.

In addition to behavior-based models for user interest, there are also incremental models that use temporal information to increase CTR. In industrial scenarios, data about user-item interactions arrive as a stream. The distribution of streaming data changes rapidly over time, and incremental models aim to capture the changing trends \cite{strengthen}. To effectively capture the latest trends, the models can be incrementally updated by taking advantage of the newly arrived data. Wang et al. \cite{practical} proposes an incremental training mode instead of batch mode. Batch mode, the common way to train deep CTR models, is trained on a window of fixed-size training data that moves forward as new data arrives. This leads to low training efficiency because training data usually overlaps on consecutive days. Incremental training captures recent trends more efficiently and is also efficient in terms of training data volume.

\subsection{Importance of Loss Functions in Deep CTR Models}
Deep CTR models are optimized by back-propagation by computing gradients to minimize the given loss on the training data. In online advertising, models are trained by minimizing the binary cross-entropy loss (BCE)  to evaluate performance through CTR. In this case, all samples are given the same optimization direction. Therefore, the temporal relationship between user-item interactions is ignored. In this work, we propose a temporal importance factor (TIF) to improve the BCE loss function by using the temporal relationship between samples.

\section{METHODOLOGY}
 \subsection{Deep CTR Models}
Deep CTR models consist of three parts: Embedding layer, interaction layer, and prediction layer.
 
\subsubsection{Embedding Layer}
CTR prediction datasets contain dense and categorical features with high cardinality. Categorical features are represented as multi-dimensional vectors using one-hot encoding. Using these sparse vectors in training has high computational cost. To avoid this, an embedding layer is applied to encode the features into low-dimensional vectors. The embedding size is a hyperparameter that specifies the size of the output vector of the embedding layer and can be adjusted during training to achieve better performance.

\subsubsection{Interaction Layer}
Various solutions have been proposed in the literature to model the feature interactions. Product transformations \cite{wideAndDeep}, multi-layer perceptron (MLP) \cite{DCN}, and factorization machine \cite{deepfm} are some examples to model second-order or higher-order interactions.

\subsubsection{Prediction Layer}
The probability of user click, $\hat{y}$, is produced in this layer. Finally, the loss function serves as the objective function using the true label and the predicted label. Typically, binary cross-entropy is used at this step.

\begin{equation}\label{BCE_eqn}
\mathcal{L}_{BCE}(y,\hat{y}) = -{(y\log(\hat{y}) + (1 - y)\log(1 - \hat{y}))}
\end{equation}

where $y$ is true label and $\hat{y}$ is predicted probability.

\subsection{Temporal Loss Factor}
The temporal importance factor was developed to account for shifts in data distribution over time. It is a weighting factor for temporal order. Chronologically ordered samples are weighted to penalize the most recent samples more. We introduce the factor in a plugin form that can be combined with various loss functions. To analyze the performance of the proposed method, we incorporated the TIF with BCE loss since it is commonly used in CTR prediction. We simply multiply the BCE loss output by the temporal importance factor, as seen in Equation \ref{temporalImp_eqn}. 

\begin{equation}\label{temporalImp_eqn}
\mathcal{L}_{BCE}(y,\hat{y}) = -{(y\log(\hat{y}) + (1 - y)\log(1 - \hat{y})) * (\alpha \frac{t}{N})}
\end{equation}

In Equation \ref{temporalImp_eqn}, $N$ is the number of days in the training dataset. $t$ is the day order of the sample in the dataset where $t \in \{1,2,...,N\}$.

Since multiplication by the temporal importance factor changes the statistical properties of the loss function, TIF can be parameterized by the constant $\alpha$. Thus, a possible destabilizing effect can be regulated. Initially, we set $\alpha$ to 1 to limit the interval for the temporal importance factor to (0,1]. To avoid a complete loss of the oldest samples, we did not include 0 in this interval.

\section{Experiments}
\subsection{Research Questions}
\begin{itemize}
\item \textbf{RQ1.} Does using the proposed temporal importance factor perform better than bce-loss?
\item \textbf{RQ2.} Are recent samples really more important for model training?
\item \textbf{RQ3.} How does the temporal weight distribution of the samples affect model training?

\end{itemize}


\subsection{Datasets}
\begin{itemize}
\item \textbf{Avazu\footnote{Avazu https://www.kaggle.com/competitions/avazu-ctr-prediction/data}:}
Avazu dataset which consists of 10 days of ad click-through data was released in a Kaggle contest . It is ordered chronologically. There are 23 fields including user, item, contextual and anonymized features.


CTR prediction models are trained using available data up to a moment to predict the future user-item dynamics. Chronological split of datasets specifies the test set as the most recent samples. In this way, experiments can reflect the real-time CTR prediction better \cite{tarik1,tarik2}.
Therefore, we split the dataset chronologically. The split ratio is 8:1:1 as training, validation, and test sets, respectively. One of the features that was provided in Avazu dataset is $hour$. Its format is $YYMMDDHH$. We followed the data pre-processing methodology as in $avazu\_x4$ of FuxiCTR \cite{fuxi}. They converted the timestamp information as $hour$, $weekday$, and $is\_weekend$.

\item \textbf{Company Dataset:}
An industrial adversiting platform CVR dataset was used for the experiments. First 7 days were selected as training set, the latter two days were used as validation and test sets, respectively. It consists of 8-days-long training data, and 1-day-long test datasets. There are 33 fields in this dataset including user features, item features, scenario-specific features, and contextual features.

\end{itemize}

\subsection{Models}
Here are the explanations of the models. DNN is a simple MLP to learn second-order feature interactions. DCN \cite{DCN}: uses cross network with DNN. DeepFM \cite{deepfm}: uses FM layer and DNN to learn second-order feature interactions and higher-order feature interactions. FiBiNET \cite{fibinet}: uses the combination of feature importance and bilinear feature interaction. PNN \cite{pnn}: has product layer after embedding layer to capture high-order interactions. MaskNet \cite{masknet} uses instance-guided mask performing element-wise product both on the feature embedding and feed-forward layers. FinalMLP \cite{finalmlp} learns feature importance and feature interactions via the Squeeze-Excitation network (SENET) mechanism and bilinear function.

\subsection{Evaluation Metrics}
\begin{itemize}
\item \textbf{Logloss:}
One of the metrics used to evaluate the results is logloss. Decreases in logloss in the level of 0.001 is considered as a significant improvement in the literature for CTR tasks \cite{autoint, adaptive}.
\end{itemize}

\begin{itemize}
\item \textbf{AUC:}
Area under curve (AUC) is a commonly used metric for CTR prediction. It should be noted that 0.1\% increase in AUC is considered as a significant improvement in CTR prediction \cite{dcn2}.

\end{itemize}

\begin{itemize}
\item \textbf{RelaImp:}
Relative improvement (RelaImp) is used to present AUC improvements better \cite{relaImp, masknet}.

\end{itemize}

\subsection{Implementation Details}
For the experiments on Avazu dataset, FuxiCTR \cite{fuxi} was used which is an open-source library for CTR prediction. It is an open benchmark including many models and different datasets. For a fair comparison, and reproducibility, we implemented the proposed solution and experimented using FuxiCTR. To show the effect of temporal importance factor, we used the experimental settings, i.e., hyperparameters, and randomness, of FuxiCTR. In this way, the performance of the solution can be claimed clearly, regardless from the hyperparameter optimization to the proposed solution. 

\section{Experimental Results}

\subsection{Performance Comparison (\textbf{RQ1})}
The effect of TIF on CTR prediction performance was evaluated using seven common and state-of-the-art deep CTR models. Training these models as typical by using BCE loss is the baseline scenario to compare with TIF. The results were shown in Table \ref{tab:temporalImp_vs_bce} using linear TIF as the most straightforward formulation to prove the impact of temporal importance factor.

\begin{table}[h]
\centering
\caption{Overall performance comparison of temporal importance factor (TIF) with bce loss (logloss and AUC) of different models on two datasets.}
\label{tab:temporalImp_vs_bce}
\begin{tabular}{cccccccc}
\Xhline{2\arrayrulewidth}
 &  & \multicolumn{3}{c}{Avazu} & \multicolumn{3}{c}{Company} \\
 \hline
 &  & logloss & AUC & RelaImp & logloss & AUC & RelaImp \\
 \hline
\multirow{2}{*}{DNN} & bce & 0.3988 & 0.7462 & \multirow{2}{*}{0.69\%} & 0.1310 & 0.7941 & \multirow{2}{*}{0.20\%} \\
 & bce+TIF & 0.3966 & 0.7479 &  & 0.1305 & 0.7947 &  \\
 \hline
\multirow{2}{*}{DCN} & bce & 0.3990 & 0.7454 & \multirow{2}{*}{0.53\%} & 0.1318 & 0.7901 & \multirow{2}{*}{0.14\%} \\
 & bce+TIF & 0.3979 & 0.7467 &  & 0.1314 & 0.7905 &  \\
 \hline
\multirow{2}{*}{DeepFM} & bce & 0.3982 & 0.7459 & \multirow{2}{*}{1.66\%} & 0.1309 & 0.7936 & \multirow{2}{*}{0.03\%} \\
 & bce+TIF & 0.3970 & 0.7500 &  & 0.1306 & 0.7937 &  \\
 \hline
\multirow{2}{*}{FiBiNET} & bce & 0.4004 & 0.7401 & \multirow{2}{*}{2.29\%} & 0.1317 & 0.7905 & \multirow{2}{*}{0.03\%} \\
 & bce+TIF & 0.3987 & 0.7456 &  & 0.1314 & 0.7906 &  \\
 \hline
\multirow{2}{*}{PNN} & bce & 0.3994 & 0.7476 & \multirow{2}{*}{0.93\%} & 0.1315 & 0.7915 & \multirow{2}{*}{0.21\%} \\
 & bce+TIF & 0.3971 & 0.7499 &  & 0.1311 & 0.7921 &  \\
 \hline
\multirow{2}{*}{MaskNet} & bce & 0.4000 & 0.7428 & \multirow{2}{*}{1.48\%} & 0.1308 & 0.7967 & \multirow{2}{*}{0.24\%} \\
 & bce+TIF & 0.3980 & 0.7464 &  & 0.1302 & 0.7974 &  \\
 \hline
\multirow{2}{*}{FinalMLP} & bce & 0.3970 & 0.7480 & \multirow{2}{*}{1.05\%} & 0.1307 & 0.7957 & \multirow{2}{*}{0.17\%} \\
 & bce+TIF & 0.3954 & 0.7506 &  & 0.1304 & 0.7962 &  \\
\Xhline{2\arrayrulewidth}
\end{tabular}
\end{table}

The results presented in Table \ref{tab:temporalImp_vs_bce} shows that TIF increases AUC and decreases logloss of all models in both datasets in the scope of the experiments. The effect is more obvious in Avazu. It is worth to emphasize that the significance of the improvement varies due to different models. The highest improvement using TIF in Avazu was in FiBiNET with 2.29\% RelaImp, while the highest increase in company dataset was achieved in MaskNet with 0.24\% RelaImp.

\subsection{Ablation Study: Anti-temporal importance factor (\textbf{RQ2})}
Target data of CTR prediction task occurs at a later time than the training in real-world uses. We hypothesize that recent samples include more beneficial information. To investigate the RQ2, we reconstructed the formulation of temporal importance factor as anti-temporal importance factor to show that focusing on older samples would decrease the model performance. It aims to function as the opposite of the idea behind temporal importance factor.

\begin{equation}\label{antitemporalImp_eqn}
\mathcal{L}_{BCE}(y,\hat{y}) = -{(y\log(\hat{y}) + (1 - y)\log(1 - \hat{y})) * (\frac{N - t + 1}{N})}
\end{equation}

This part of the study was conducted to better understand the significance of the temporal importance phenomena that we focus on. DNN was used as the base model. Table \ref{tab:anti_TemporalImp} shows that learning more from older samples decreases the performance of DNN. Looking from this opposite perspective of the proposed method, it makes sense to train the models by focusing more on the recent samples.

\begin{table}[H]
\centering
\caption{Ablation study using anti-TIF. Anti-TIF forces the model to focus on the older samples.}
\label{tab:anti_TemporalImp}
\begin{tabular}{cccc}
\Xhline{2\arrayrulewidth}
 &  & logloss & AUC \\
 \hline
\multirow{2}{*}{DNN} & \textbf{bce} & \textbf{0.39887} & \textbf{0.7463} \\
 & bce+anti-TIF & 0.40010 & 0.7445 \\
\Xhline{2\arrayrulewidth}
\end{tabular}
\end{table}

\subsection{Investigating the effect of weight distribution by temporal importance factor (\textbf{RQ3})}

We suggest that temporal order of the samples is important for model training. Moreover, the distribution of the temporal importance is another question. In our first simple proposal of the temporal importance factor, there was a linear weighting of the samples. However, there can be nonlinear distribution of the temporal importance. To investigate this question, we constructed different formulations for the temporal importance factor. It should be noted that this distribution can vary due to different datasets. Therefore, different formulations can be experimented and our general solution for temporal importances in CTR prediction data can be optimized for different industries.

\begin{figure}[h]
  \centering
  \includegraphics[scale=0.3]{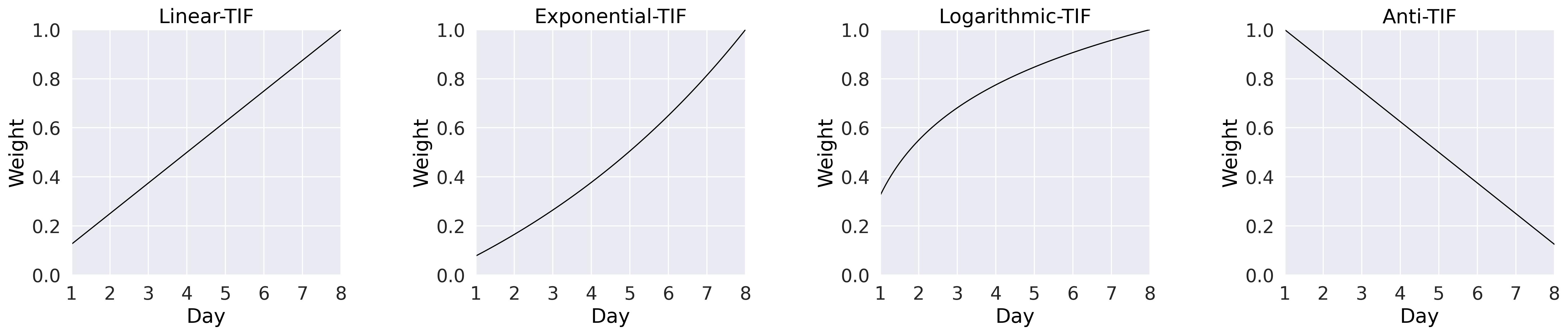}
  \caption{TIF variants that weight samples by day information are shown as Linear-TIF, Exponential-TIF, Logarithmic-TIF, and Anti-TIF respectively.}
  \label{FormulGraphs}
\end{figure}

In addition to basic linear TIF, we further proposed two variations of temporal importance factor: (\emph{i}) exponential, and (\emph{ii}) logarithmic. The aim of this selection is to observe the temporal weight distribution changes by shifting the mean weight to right and left, respectively. The formulations are shown in Equation \ref{EXPtemporalImp_eqn} and \ref{LOGtemporalImp_eqn}, respectively. The different variants are illustrated in Figure \ref{FormulGraphs}.

\begin{equation}\label{EXPtemporalImp_eqn}
\mathcal{L}_{BCE}(y,\hat{y}) = -{(y\log(\hat{y}) + (1 - y)\log(1 - \hat{y})) * (\frac{e^{t} - 1}{e^N - 1})}
\end{equation}

\begin{equation}\label{LOGtemporalImp_eqn}
\mathcal{L}_{BCE}(y,\hat{y}) = -{(y\log(\hat{y}) + (1 - y)\log(1 - \hat{y})) * (\frac{log(t) + 1}{log(N) + 1})}
\end{equation}

The results are presented in Table \ref{tab:temporalImp_versions}. All of the variants perform better than baseline. Exponential-TIF mostly performs the best. Since it focuses much more on recent samples than linear- and logarithmic-TIF variants, this can indicate that the proposed methods. 

\begin{table}[]
\centering
\caption{Comparison of linear-, exponential-, and logarithmic-TIF on Avazu dataset.}
\label{tab:temporalImp_versions}
\begin{tabular}{ccccc}
\Xhline{2\arrayrulewidth}
 &  & logloss & AUC & RelaImp \\
 \hline
\multirow{3}{*}{DNN} & linear & 0.3966 & 0.7480 & 0.73\% \\
 & logarithmic & 0.3978 & 0.7469 & 0.28\% \\
 & exponential & \textbf{0.3961} & \textbf{0.7489} & \textbf{1.10\%} \\
 \hline
\multirow{3}{*}{DCN} & linear & 0.3979 & 0.7467 & 0.53\% \\
 & logarithmic & 0.3990 & 0.7463 & 0.37\% \\
 & exponential & \textbf{0.3970} & \textbf{0.7479} & \textbf{1.02\%} \\
 \hline
\multirow{3}{*}{DeepFM} & linear & 0.3971 & 0.7500 & 1.67\% \\
 & logarithmic & 0.3975 & 0.7490 & 1.26\% \\
 & exponential & \textbf{0.3962} & \textbf{0.7508} & \textbf{1.99\%} \\
 \hline
\multirow{3}{*}{FiBiNET} & linear & \textbf{0.3987} & \textbf{0.7457} & \textbf{2.33\%} \\
 & logarithmic & 0.3998 & 0.7427 & 1.08\% \\
 & exponential & 0.4001 & 0.7410 & 0.37\% \\
 \hline
\multirow{3}{*}{PNN} & linear & 0.3971 & 0.7480 & 0.16\% \\
 & logarithmic & \textbf{0.3971} & \textbf{0.7497} & \textbf{0.85\%} \\
 & exponential & 0.3973 & 0.7493 & 0.69\% \\
 \hline
\multirow{3}{*}{MaskNet} & linear & 0.3981 & 0.7464 & 1.48\% \\
 & logarithmic & 0.3981 & 0.7465 & 1.52\% \\
 & exponential & \textbf{0.3976} & \textbf{0.7467} & \textbf{1.61\%} \\
 \hline
\multirow{3}{*}{FinalMLP} & linear & 0.3955 & 0.7507 & 1.09\% \\
 & logarithmic & 0.3964 & 0.7496 & 0.65\% \\
 & exponential & \textbf{0.3955} & \textbf{0.7511} & \textbf{1.25\%} \\
\Xhline{2\arrayrulewidth}
\end{tabular}
\end{table}

\section{Conclusion}
In this paper, we first introduce a problem related to the time-dependent nature of CTR data, i.e., data distribution shift over time, concept drift. This hinders the ability of deep CTR models to capture recent trends. To overcome this problem, we proposed the temporal importance factor for loss functions. TIF forces the model to learn more from the recent samples by penalizing the loss values of them more. We used TIF along with BCE, thus we use BCE only as the baseline for comparisons. TIF improved the performance of common and state-of-the-art models on both Avazu and company datasets. The formulation of TIF can be adapted to the varying natures of different application areas of CTR prediction. We experimented with three different variants of TIF: (i) linear, (ii) logarithmic, and (iii) exponential. The result was that exponential-TIF produced the best improvements. Since it focuses much more on recent samples than other TIF variants, these results also support our approach to the problem. Beyond using BCE, TIF can also be used with other loss functions to understand its behavior with them. It can be applied to areas that have a time-dependent nature to further investigate the effects.

\bibliographystyle{unsrt}  
\bibliography{references}

\end{document}